\documentstyle[12pt,psfig,a4]{article}
\begin{document}
\def\doublespaced{\baselineskip=\normalbaselineskip\multiply\baselineskip
  by 150\divide\baselineskip by 100}
\doublespaced
\def\lsim{~{\rlap{\lower 3.5pt\hbox{$\mathchar\sim$}}\raise 1pt\hbox{$<$}}\,}
\def\gsim{~{\rlap{\lower 3.5pt\hbox{$\mathchar\sim$}}\raise 1pt\hbox{$>$}}\,}
\def\thisday{~\today ~and~ gr-qc/0004027~~}

\begin{titlepage}
\vspace{0.5cm}
\begin{flushright}
April 2000
\end{flushright}
\vspace{0.5cm}
\begin{center}
\large
{Boson Condensation in an Einstein Universe}
\end{center}
\begin{center}
{\bf M.B. Altaie$^a$\footnote{
e-mail:maltaie@yu.edu.jo}} and {\bf Ehab Malkawi$^b$}\\
$^a${Department of Physics, University of Yarmouk\\
Irbid 21163, Jordan}\\
$^b${Department of Physics,
Jordan University of Science \& Technology\\
 Irbid 22110, Jordan}
\end{center}
\vspace{0.4cm}
\raggedbottom
\relax
\begin{abstract}
\noindent
In this paper we investigate the Bose-Einstein condensation of massive 
spin-1 particles in an Einstein universe. The system is considered under 
relativistic conditions taking into 
consideration the possibility of particle-antiparticle pair production. 
An exact expression for the charge density is obtained, then certain 
approximations are employed in order to 
obtain the solutions in closed form. A discussion of the approximations 
employed in this and other work is given. The effects of finite-size and 
spin-curvature coupling are emphasized.   
\end{abstract}
\end{titlepage}

\section{ Introduction}
Since the early work of Altaie (1978), the study of Bose-Einstein condensation 
(BEC) in curved space 
has attracted the interest of many authors (Carvalho and Rosa 1980, Singh and 
Pathria 1984, Parker and Zhang 1991, Smith and Toms 1996, and Trucks 1998). 
This interest comes in the context 
of trying to understand the thermodynamics of the early universe and the role 
played by the finiteness of the space in determining the thermal behavior of 
bosons in the respective systems.
 
Most studies employ the static Einstein universe as the underlying geometry 
for the system, since 
the thermodynamic equilibrium in this universe can be defined without ambiguity 
(Altaie and Dowker 1978). 
In a time developing spacetime no such luxury is enjoyed, except for the 
Robertson-Walker spacetime which is conformally static (Kennedy 1978).

In an earlier work (Altaie 1978), we considered the cases of non-relativistic 
BEC of spin 0 and spin 1 bosons in an Einstein universe. We  found that the 
finiteness of the system resulted in 
"smoothing out" the singularities of the thermodynamic functions found in 
the infinite systems, 
the enhancement of the condensate fraction and the displacement of the 
specific-heat maximum toward higher temperatures.
        
Carvalho and Rosa (1980) considered the case of relativistic scalar field in 
an Einstein universe in an effort to show that the finite size effects are 
negligible in comparison with the relativistic 
effects. However, a close look at their work shows that this comparison was 
done at implicitly large value of the radius $a$ of the Einstein universe. 
And since finite size effects are proportional to $1/a$ the finite size 
effects were found to be negligible. Moreover, the consideration of relativistic 
and ultra-relativistic BEC should take into account the possibility of 
particle-antiparticle 
production, since at temperatures greater than the rest mass of the particles, 
quantum field theory 
requires the inclusion of such a process (Harber and Weldon 1981, 1982).   

Singh and Pathria (1984) considered the BEC of a relativistic  conformally 
coupled scalar field in 
the Einstein universe and found qualitative and quantitative agreement 
with Altaie (1978). Parker 
and Zhang (1991) considered the ultra-relativistic BEC of the minimally coupled 
scalar field in an 
Einstein universe in the limit of high temperatures. They showed, among other 
things that ultra-relativistic BEC can occur at very 
high temperature and 
densities in the Einstein universe, and 
by implication in the early stages of a dynamically changing universe. 
Parker and Zhang (1993) also showed that the Bose -Einstein condensate could act as 
a source for inflation leading to a de Sitter type universe.

 Trucks (1998) repeated the calculations of Singh and Pathria but this time 
for the minimally coupled scalar field obtaining similar results. In fact 
performing the calculations for the 
minimally coupled scalar field amounts to substituting 
$\overline{m}={(m^2+1/a^2)}^{1/2}$ where $m$ is the mass 
of the conformally coupled field. But as the calculations were effectively 
considered in the large radius region, 
the results come out to be identical with those of Singh and Pathria.

The importance of the study of BEC in curved spaces stems from the interest 
in understanding the thermodynamics of the very early universe and that 
such a phenomenon may shed some light on 
the problem of mass generation in the very early universe. Indeed the 
investigations of Toms (1992, 1993) have shown that a kind of symmetry 
breaking is possible. 

In this paper we will 
consider the onset of BEC of the relativistic spin 1 particle-antiparticle 
system in an Einstein universe, a state which is surly relevant for the early 
stages of the universe at a point when 
the electromagnetic interactions decouple from the weak interactions. We will 
carry out the calculations in a similar fashion to that of Singh and Pathria 
and compare the results with our earlier non-relativistic case. Throughout this 
work we adopt the absolute system of units in which $c=G=k=\hbar=1$. 

\section{The Charge Density}

We consider an ideal relativistic Bose gas of spin 1 confined to the background 
geometry of the spatial section $S^3$ of an Einstein universe with  radius $a$. 
Since we are considering a relativistic 
system, it is necessary to take in consideration the possibility of pair 
production. The system is taken to be formed of $N_1$ particles and $N_2$ 
antiparticles. The total number of particles $Q=N_1-N_2$ is assumed 
to be conserved, though  $N_1$ and $N_2$ may change. A chemical potential 
$\mu$ (is assigned for the particles 
and $-\mu$ for the antiparticles (see Harber and Weldon 1981). 
Accordingly the particle and antiparticle distributions are given by,
\begin{equation}
N_1=\sum_{n} d_n {\left[ e^{\beta \left(\epsilon_n-\mu\right)}-1\right]}^{-1}\, ,
\hspace{1cm} 
N_2=\sum_{n} d_n {\left[ e^{ \beta \left(\epsilon_n+\mu\right)}-1\right]}^{-1}\, ,
\label{N}
\end{equation}
where $\beta=1/T$, $\epsilon_n$ are the eigen energies, and $d_n$ is the degeneracy 
of the $n$th level. 
The equation of motion of the spin 1 field in an Einstein universe was considered 
by  Schr\"{o}dinger (1938) and the solution yields the following energy spectrum, 
\begin{equation}
\epsilon_n=\frac{1}{a} {\left(n^2+m^2a^2\right)}^{1/2}\, ,
\end{equation}
with degeneracy,
\begin{equation}
d_n=2(n^2-1)\, , \hspace{1cm} n=2,3,4,...
\end{equation}
The charge density $q$ is then found to be,
\begin{equation}
q=\frac{Q}{V}=\frac{4}{V}\sum_{l=1}^{\infty}\sum_{n=1}^\infty (n^2-1)\sinh(l\beta\mu)
\exp\left[-l\beta^\prime{\left(m^2a^2+n^2\right)}^{1/2}\right]\, ,
\label{q1}
\end{equation}
where $V=2\pi^2 a^3$ is the volume of the spatial section of the Einstein universe, and 
$\beta^\prime=1/Ta$. In order to carry out the summation over $n$ in (\ref{q1}) we apply 
the Poisson summation formula (see Titchmarsh 1948),
\begin{equation}
\sum_{n=1}^\infty f(n)+ \frac{1}{2}f(0)=\int_{0}^\infty f(t)dt 
+2\sum_{j=1}^\infty \int_0^\infty f(t)\cos(2\pi jt) dt\, .
\label{poisson}
\end{equation}
Accordingly,
\begin{equation}
\sum_{n=1}^\infty (n^2-1)
\exp\left[-l\beta^\prime{\left(m^2a^2+n^2\right)}^{1/2}\right]=
\frac{1}{2} \exp(l\beta m) +I_0 +2\sum_{j}^\infty I_j\, ,
\end{equation}
where,
\begin{equation}
I_0=\int_0^\infty (t^2-1)\exp\left[-l\beta^\prime
{\left(m^2a^2+t^2\right)}^{1/2}\right]dt\, ,
\end{equation}
and
\begin{equation}
I_j=\int_0^\infty (t^2-1)\exp\left[-l\beta^\prime
{\left(m^2a^2+t^2\right)}^{1/2}\right]\cos(2\pi jt)dt\, .
\end{equation}
These integrals can be easily performed using (Gradshteyn and Ryzhik 1980)
\begin{equation}
\int_0^\infty \exp\left[-\alpha\sqrt{\gamma^2+x^2}\right]\cos(\lambda x)dx=
\frac{\gamma\alpha}{\sqrt{\lambda^2+\alpha^2}} K_1
\left(\gamma\sqrt{\lambda^2+\alpha^2}\right)\, ,
\end{equation}
where $K_\nu$ are the modified Bessel functions of the second kind.

Evaluating $I_0$ and  $I_j$ and then substituting in (\ref{q1}) we find that the 
charge density can be written as,
\begin{eqnarray}
&&q=\frac{1}{2\pi^2a^3}\left[\frac{1}{e^{\beta(m-\mu)}-1}-
\frac{1}{e^{\beta(m+\mu)}-1}\right]\cr
&&-\frac{2m}{\pi^2a^2}\sum_{l=1}^{\infty}\sum_{j=1}^\infty l \sinh(l\beta\mu)
\left[\frac{K_1(l \beta m)}{l} +\frac{K_1(\beta m z)}{z}\right]\cr
&&+\frac{2m^2}{\pi^2\beta}\sum_{l=1}^\infty\sum_{j=-\infty}^\infty l\sinh(l \beta\mu )
\left[\frac{K_2(\beta m z)}{z^2}-(\beta m j^\prime)^2\frac{K_3(\beta mz)}{z^3}\right]\, ,
\label{q2}
\end{eqnarray}
where 
\begin{equation}
z=\sqrt{l^2+{j^\prime}^2}\,\,\, \, {\rm{and}}\,\,\,\,  j^\prime=(2\pi a/\beta)j\, .
\end{equation}
 The first term in (\ref{q2}) arises because the $n=0$ term in the summation 
over $n$ in (\ref{q1}) is non-zero. The second term arises because of the spin-curvature 
coupling, which is defined mathematically by the coefficient $a_n$ of the 
Schwinger-De Witt expansion (see, De Witt 1965). The third term is just twice 
that of the scalar case considered by Singh and Pathria (1984). 
Both the first and the second terms disappear in the limit $a\rightarrow \infty$. 
The bulk term is obtained in this limit assuming that $q$ remains constant, 
which is the known thermodynamic limit. This reduces to the $j=0$ contribution of  
the last term in (\ref{q2}) which gives,
\begin{equation}
q_B(\beta,\mu)=\frac{2m^3}{\pi^2}\sum_{l=1}^\infty {(l\beta m)}^{-1} 
\sinh(l \beta \mu ) K_2(l \beta m)\, .
\end{equation}
This is just twice the value obtained for the scalar case as would be expected.
The summation over $j$ in (\ref{q2}) can be performed using the Poisson summation 
formula (\ref{poisson}) again, where we obtain, 
\begin{eqnarray}
&&q=q_B(\beta,\mu)+ \left[\frac{1}{e^{\beta(m-\mu)}-1}-
\frac{1}{e^{\beta(m+\mu)}-1}\right]\cr
&&-\frac{2m}{\pi^2a^2}\sum_{l=1}^{\infty}\sum_{p=1}^\infty \int_0^\infty 
l \sinh(l\beta\mu^\prime)
\left[\frac{K_1(l\beta m)}{l} +\frac{K_1(\beta m z)}{z}\right] dj \cr
&&+\frac{4m^2}{\pi^2\beta}\sum_{l=1}^\infty\sum_{p=1}^\infty \int_0^\infty 
l\sinh(l \beta\mu^\prime )
\left[\frac{K_2(\beta m z)}{z^2}-(\beta mj)^2\frac{K_3(\beta mz)}{z^3}\right] dj\, ,
\end{eqnarray}
where $
\mu^\prime =\mu +2\pi i p$.
The integrals in the above equation can be evaluated exactly using 
(see Singh and Pathria 1984),
\begin{equation}
\int_0^\infty j\sinh(jx)\frac{K_\nu\left[y{(j^2+\xi^2)}^{1/2}\right]}{{(j^2+\xi^2)}^{\nu/2}}dj=
{\left(\frac{\pi\xi^3}{2}\right)}^{1/2}\frac{x{(y^2-x^2)}^{\nu/2-3/4}}{(\xi y)^\nu}
K_{\nu-3/2}\left[\xi(y^2-x^2)^{1/2}\right]\, .
\end{equation}
Then using (see Gradshteyn and Ryzhik 1980),
\begin{equation}
\int_0^\infty \sinh(ax)K_1(bx)dx= \frac{\pi}{2} \frac{a}{b\sqrt{b^2-a^2}}\, ,
\end{equation}
and the relation,
\begin{equation}
K_{1/2}(z)-zK_{3/2}(z)=-zK_{-1/2}(z)=-{\left(\frac{\pi z}{2}\right)}^{1/2} e^{-z}\, ,
\end{equation}
we obtain
\begin{eqnarray}
&&q=q_B(\beta,\mu)+\frac{1}{2\pi^2 a^3} \left[\frac{1}{e^{\beta(m-\mu)}-1}-
\frac{1}{e^{\beta(m+\mu)}-1}\right]\cr
&&-\frac{2}{\pi\beta}\sum_{l=1}^\infty \sum_{p=-\infty}^{\infty} \mu^\prime
\left[\frac{1}{2a^2\tau}+\left(\frac{1}{a^2\tau}+\tau\right) e^{-2\pi al\tau}\right]\, ,
\end{eqnarray}
where $\tau=\sqrt{m^2-{\mu^\prime}^2}$.

The summation over $l$ can be easily done, so the exact expression for the 
charge density becomes,
\begin{eqnarray}
&&q=q_B(\beta,\mu)+
\frac{1}{2\pi^2 a^3} \left[\frac{1}{e^{\beta(m-\mu)}-1}-
\frac{1}{e^{\beta(m+\mu)}-1}\right]\cr
&&-\frac{2}{\pi\beta}\sum_{p=-\infty}^{\infty} \mu^\prime
\left[\frac{1}{2a^2\tau}+\left(\frac{1}{a^2\tau}+\tau\right)
\frac{1}{e^{2\pi a\tau}-1}\right]\, .
\label{q3}
\end{eqnarray}
Note that this form of the charge density is exact and no approximation 
whatsoever has been made through the calculation.

\section{Bose-Einstein Condensation}

We will adopt the microscopic criteria for marking the onset of the condensation 
(Altaie 1978), according to which the condensation region is defined such that a 
large number of particles is found occupying the ground state. This implies that 
the chemical potential $\mu$ of the system approaches the minimum single particle 
energy $\epsilon_2$, not the single rest mass energy considered by 
Singh and Pathria (1984). 
This can be observed directly from (\ref{N}). However the consideration of the criteria 
that $\mu\rightarrow m$ is justified only for the minimally coupled scalar field
case where the 
minimum energy is $m$ (see Parker and Zhang 1991). In our case the chemical potential 
must satisfy the condition that
\begin{equation}
-{\left(m^2+\frac{4}{a^2}\right)}^{1/2} < \mu < {\left(m^2+\frac{4}{a^2}\right)}^{1/2}\, ,
\end{equation}
whereas in the conformally coupled spin 0 case the condition is,
\begin{equation}
-{\left(m^2+\frac{1}{a^2}\right)}^{1/2} < \mu < {\left(m^2+\frac{1}{a^2}\right)}^{1/2}\, .
\end{equation}
However, if $m^2$ is much larger than $1/a^{2}$  then it will be justified to 
take the limit $\mu\rightarrow m$, but this approximation may impose certain 
restrictions  on the range of the region under consideration. We will follow 
Singh and Pathria and adopt such approximation in this work. In such a case 
the main contribution of the summation over $p$ in (\ref{q3}) comes from the $p=0$ 
term. Other terms are of order of $e^{-a\sqrt{m/\beta}}$, i.e. $O(e^{-a/\lambda_T})$ 
where $\lambda_T=\sqrt{2\pi\beta/m}$ is the mean thermal wavelength of the particle.
The bulk term will reduce to (Singh and Pandita 1983),
\begin{equation}
q_B(\beta,\mu)=q_B(\beta,m)-\frac{m}{\pi \beta}{\left(m^2-\mu^2\right)}^{1/2}+
O(m^2-\mu^2)\, .
\end{equation}
Therefore we can write
\begin{eqnarray}
&&q_B(\beta,\mu)\approx q_B(\beta,m)-\frac{m}{\pi \beta}{\left(m^2-\mu^2\right)}^{1/2}\cr
&&-\frac{2\mu}{\pi\beta}\left[{(m^2-\mu^2)}^{1/2}-\frac{1}{a^2}{(m^2-\mu^2)}^{-1/2}\right]
\frac{1}{\exp{\left(2\pi a \sqrt{m^2-\mu^2}\right)}-1}\cr
&& +\frac{1}{2\pi^2 a^3}\left[\frac{1}{e^{\beta(m-\mu)}-1}-
\frac{1}{e^{\beta(m+\mu)}-1}\right]\,
\label{q4} .
\end{eqnarray}
If we define the thermogeometric parameter y as 
\begin{equation}
y=\pi a {(m^2-\mu^2)}^{1/2}\, ,
\label{q6}
\end{equation}
then equation (\ref{q4}) becomes
\begin{equation}
q\approx q_B(\beta,m)-\frac{m}{\pi \beta a}\left[ \left(\frac{y}{\pi}+ \frac{\pi}{y}\right)\coth y
-\frac{\pi}{y^2}\right]\,
\label{q5} .
\end{equation}
From this equation the behavior of the thermogeometric parameter $y$ in the 
condensation region can be determined. It is clear that the second term in (\ref{q5}) 
defines the finite-size and spin-curvature effects.

\section{The Condensate Fraction} 

The growth of the condensate fraction is studied here in comparison 
with the bulk case. This will show the finite-size and the spin-curvature 
effect in the range considered, i.e.,   $ma\gg 1$.
The charge density in the ground state is obtained if we substitute $n=2$ 
in (\ref{N}). This gives 
\begin{equation}
q_0=\frac{6}{2\pi^2 a^3}\left[ {\left( e^{\beta(\epsilon_2-\mu)}-1\right)}^{-1}- 
{\left(e^{\beta(\epsilon_2+\mu)}-1\right)}^{-1}\right] \, .
\end{equation}

In the condensation region $\mu \rightarrow \epsilon_2$ , so that the main 
contribution to the charge density in the ground  state comes from the first 
term in the square bracket. This means that
\begin{equation}
q_0\approx \frac{3}{\pi^2 a^3\beta (\epsilon_2-\mu)}\, .
\label{q7}
\end{equation}
From (\ref{q6}) as $\mu\rightarrow m$ , we have
\begin{equation}
\mu\approx m\left( 1-\frac{y^2}{2\pi^2 a^2 m^2}\right)\, .
\end{equation}
On the same footing we can expand $\epsilon_2$ as
\begin{equation}
\epsilon_2\approx m\left(1+\frac{2}{m^2 a^2}\right)\, .
\end{equation}
So that (\ref{q7}) becomes
\begin{equation}
q_0\approx \frac{6m}{a\beta (y^2+4\pi^2)}\,.
\label{q8}
\end{equation}
This means that the macroscopic growth of the condensate will occur only 
when $y^2\rightarrow -4\pi^2$, (i.e., $y\rightarrow 2\pi i$). 
In order to see how this condensate compares with the bulk case we use 
the expansion:
\begin{equation}
\coth y=\frac{1}{y}+2y\sum_{k=1}^\infty \frac{1}{y^2+\pi^2 k^2}\, .
\end{equation} 
So that,
\begin{equation}
\left(\frac{y}{\pi}+\frac{\pi}{y}\right)\coth y -\frac{\pi}{y^2}=\frac{5}{\pi}
-\frac{6\pi}{y^2+4\pi^2}+\frac{2}{\pi}(y^2+\pi^2)\sum_{k=3}^\infty\frac{1}{y^2+\pi^2 k^2}\, .
\end{equation}
Substituting this in (\ref{q5}) and using (\ref{q8}) we get,
\begin{equation}
q_0=q-q_B(\beta,m)+\frac{2m}{\pi^2\beta a}\left[ \frac{5}{2}+(y^2+\pi^2)\sum_{k=3}^\infty
\frac{1}{y^2+\pi^2 k^2}\right]\,.
\end{equation}
For the bulk system ($a\rightarrow \infty$) there exists a critical temperature, $T=T_c$
given by 
\begin{equation}
q_B(\beta_c,m)=q\, .
\end{equation} 
This condition can be written as 
\begin{equation}
q=q_B(\beta_c,m)=\frac{m^3}{\pi^2}\sum_{l=1}^\infty {(l\beta_c m)}^{-1}
\sinh(l\beta_c m) K_2(l\beta_c m)=\frac{m^3}{2\pi^2}W(\beta_c,m)\, ,
\label{q13}
\end{equation}
where
\begin{equation}
W(\beta,\mu)=2\sum_{l=1}^\infty {(l\beta\mu)}^{-1} \sinh(l\beta\mu)K_2(l\beta\mu)\, .
\end{equation}
Thus we can write the bulk condensate density as
\begin{equation}
(q_0)_B=\left\{ \begin{array}{ll}
0 &\,\, \mbox{for \,\,$T>T_c$}\\
q\left(1-\frac{W(\beta,m)}{W(\beta_c,m)}\right) & \,\,\mbox{for \,\,$T<T_c$}
\end{array}\right. \, .
\label{q11}
\end{equation}
For the case of the finite system under consideration the condensate density is given by
\begin{equation}
q_0=q\left[1-\frac{W(\beta,m)}{W(\beta_c,m)}\right] +
\frac{2m}{\pi^2\beta a}\left[\frac{5}{2}+(y^2+\pi^2)\sum_{k=3}^\infty\frac{1}{y^2+\pi^2 k^2}\right]\, ,
\end{equation}
where the $y$ dependence on $T$ in the condensation region
 can now be determined explicitly by 
\begin{equation}
\left(\frac{y}{\pi}+\frac{\pi}{y}\right)\coth y -\frac{\pi}{y^2}=
-\frac{\pi \beta aq}{m}\left[ 1-\frac{W(\beta,m)}{W(\beta_c,m)}\right]\, .
\label{q10}
\end{equation}
This equation can be solved numerically for the values of $y$ at different 
temperatures. However, we notice that at $T=T_c$  
$\left( W(\beta,m)=W(\beta_c,m)\right)$, the value of $y$ can be obtained 
by solving the equation,
\begin{equation}
\left(\frac{y_c}{\pi}+\frac{\pi}{y_c}\right)\coth y_c =\frac{\pi}{y_c^2}\, ,
\end{equation} 
which has a solution at $y_c=4.859i$.

\section{Non-relativistic and Ultra-relativistic Limits}

The non-relativistic limit is obtained by setting $\beta m\gg 1$. In this case 
we can use the asymptotic expansion of the Bessel functions of the second kind 
for large argument, where we have (see Abramowitz and Stegun 1968), 
\begin{equation}
K_2(j \beta m)\approx {\left(\frac{\pi}{2j\beta m}\right)}^{1/2} e^{-j\beta m}
\left[ 1+\frac{15}{8}\frac{1}{j\beta m}+\frac{105}{128}\frac{1}{{(j\beta m)}^2}+...\right]\, .
\end{equation}
So that
\begin{equation}
\left[ W(\beta,m)\right]_{NR}={\left(\frac{\pi}{2m^3}\right)}^{1/2}\zeta(3/2)\beta^{-3/2}
\end{equation}
and
\begin{equation}
\left[\frac{W(\beta,m)}{W(\beta_c,m)}\right]_{NR}={\left(\frac{\beta_c}{\beta}\right)}^{3/2}
={\left(\frac{T}{T_c}\right)}^{3/2}\, .
\end{equation}
Therefore from (\ref{q11}) we have for the bulk system,
\begin{equation}
(q_0)_B=q\left[1-{\left(\frac{T}{T_c}\right)}^{3/2}\right]\, ,\,\,\, T<T_c
\end{equation}
In the non-relativistic limit (\ref{q10}) reduces to
\begin{equation}
2\left(\frac{y}{\pi}+\frac{\pi}{y}\right)\coth y -\frac{2\pi}{y^2}=
x^{1/2}\left[1-x^{-3/2}\right]\frac{a}{\bar{l}}{\left[2\zeta(3/2)\right]}^{2/3}\, ,
\label{q9}
\end{equation}
where $x=T/T_c$ and $\bar{l}$ is the mean inter-particle distance. 
Note that equation (\ref{q9}) is the corrected version of equation (62) 
of our earlier paper (Altaie 1978).  

In the ultra-relativistic limit $\beta m\ll 1$, therefore, we use the expansion,
\begin{equation}
K_2(j\beta m)\sim \frac{1}{2}\Gamma(2){\left(\frac{1}{2}j\beta m\right)}^{-2}\, .
\end{equation}
So that,
\begin{equation}
\left[W(\beta,m)\right]_{UR}\sim \frac{2\pi^2}{3m^2}\beta^{-2}\, .
\label{q12}
\end{equation}
Accordingly the ultra-relativistic behavior of the bulk charge density 
given in (\ref{q11}) 
will be 
\begin{equation}
(q_0)_B=q\left[1-\frac{T^2}{T_c^2}\right]\, .
\end{equation}
However, substituting (\ref{q12}) in (\ref{q13}), 
the critical temperature for the bulk spin 1 particles is
\begin{equation}
T_c=\sqrt{\frac{3q}{2m}}\, .
\label{q14}
\end{equation}   
This is the analogous result to that of the minimally coupled scalar field.
This shows that the treatment of the problem using the large radius 
approximation as adopted by Singh and Pathria and in this work is equivalent 
to the high temperature expansion of Parker and Zhang (1991). This can be 
understood in the light of the fact that in the ultra-relativistic regime 
$T\propto a^{-1}$, so that $aT=$constant.
In the ultra-relativistic regime the thermogeometric parameter $y$ behaves
according to
\begin{equation}
\left( \frac{y}{\pi}+\frac{\pi}{y}\right) \coth y=
\frac{\pi aq}{mT_c}\left(x-\frac{1}{x}\right)=
\pi a \sqrt{\frac{2q}{3m}}\left(x-\frac{1}{x}\right)\, .
\label{q15}
\end{equation}
\begin{figure}[t]
\begin{center}
\leavevmode\psfig{file=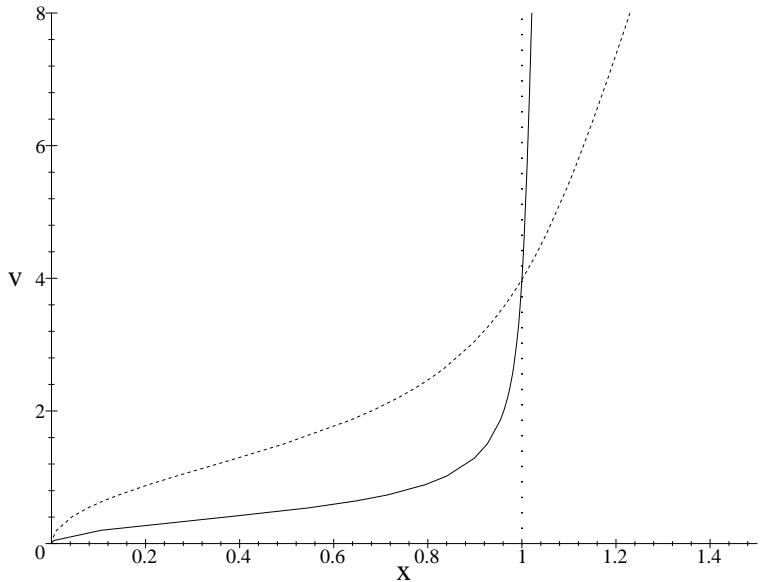,angle=270,height=10.cm}
\end{center}
\caption{The quantity $v=\sqrt{y^2+4\pi^2}$, where $y$ is the 
thermogeometric parameter, is plotted as a function of the scaled
temperature $x=T/T_c$ for two different values of the
radius $a$. The solid line is for the larger value of $a$. The vertical
dotted line at $x=1$ is marked for reference.}
\label{fig1}
\end{figure}
This equation can be solved for given values of $a$, $q$, and $m$.
For illustrative purposes, figure 1 shows the behavior of the thermogeometric parameter 
$y$ (drawn as $v=\sqrt{y^2+4\pi^2}$ versus the scale temperature 
$x=T/T_c$).
It is clear that as $a$ increases the quantity $\sqrt{y^2+4\pi^2}$ tends 
to a step function.

\section{The Critical Radius}

Although it is known that the universe as a whole is neutral or almost neutral, 
the observational limit on the net average charge density do not exclude the 
possibility that the charge density in the early universe was sufficient to 
produce relativistic Bose-Einstein condensation. Following the assumption 
that the charge is conserved we can write
\begin{equation}
q_i={\left(\frac{a_p}{a_i}\right)}^3 q_p\, ,
\end{equation}
where $q_i$ and $q_p$ are the initial and the present charge densities, respectively.
If we consider $Ta=C$ and assume that $C$ remains constant throughout the development 
of the universe then from (\ref{q14}) we can allocate a critical radius for the universe $a_c$ 
below which the gas will be always in the condensate state. Parker and Zhang (1991) 
have already noted this. However for the spin 1 case this critical radius will be given by
\begin{equation}
a_c=\frac{3q_p a_p^3}{2mC^2}\, .
\end{equation}
The estimated upper bound on the net average charge density of the universe at present 
is $q<10^{-24}$ cm$^{-3}$ (see Dolgov and Zeldovich 1981). If this upper bound on the 
present charge density is adopted, then an upper bound on the radius of the universe 
at which condensate starts can be deuced. Using a value of $10^{28}$ for $C$ 
(Trucks 1998) and 
$a_p =10^{28}$ cm, the critical radius for the onset of the condensation 
of the heavy gauge bosons $W^{±}$ can be calculated. This gives $a_c <10^{-11}$ cm.

\section{Discussion}
In the previous sections we have investigated the behavior of an ideal relativistic 
spin-1 gas confined to the background geometry of an Einstein universe. 
The extension of the problem into the Robertson-Walker spacetime is straight 
forward and results will be similar to the ones obtained here apart from the 
fact that in the Robertson-Walker spacetime the radius $a$ is time-dependent, 
so $a\rightarrow a(t)$. On the other hand more accurate calculations may be 
needed in order to understand the behavior of bosons at very early stages 
of the development of the universe. In this context the problem of photon 
condensation will surely  be of great interest, however some technical and 
conceptual problems hinders obtaining exact results.  

\section*{Acknowledgment} 
One of the authors M.B.A. would like to thank R.K. Pathria for useful discussions.

\newpage

\section*{References}
Abramowitz M and Stegun I A 1968 {\it Handbook of Mathematical Functions} 
(new York:dover)\\
Altaie M B 1978 {\it J. Phys. A: Math. Gen. 11 1603}\\
Altaie M B and Dowker J S 1978 {\it Phys. Rev. D 18 3557}\\
Aragao de Carvalho C A and Goulart Rosa S Jr 1980 {\it
J. Phys. A: Math. Gen. 13 989}\\
De Wit B S 1965 {\it Dynamical Theory of Groups and Fields} (New York: Gordon and Beach)\\
Dolgov A and Zeldovich Y 1981 {\it Rev. Mod. Phys. 53 1}\\
Gradshteyn I S and Ryzhik I M 1980 {\it Tables of Integrals, Series and Products} 
(New York: Academic Press) \\
Haber H E and Weldon H A 1981 {\it Phys. Rev. Lett. 46 1497}\\
{---------} 1982 {\it Phys. Rev. D 25 502}\\
Kennedy G 1978 {\it J. Phys. A 11 L77}\\
Parker L and Zhang Y 1991 {\it Phys. Rev. D 44 2421}\\
{---------} 1993 {\it Phys. Rev. D 47 2483}\\
Schr\"{o}dinger E 1938 {\it Comm. Pont. Acad. Sci. 2 321}\\
Singh S and Pandita P N 1983 {\it Phys. Rev. A 28 1752}\\
Singh S and Pathria R K 1984 {\it J. Phys. A: Math. Gen. 17 2983}\\
Smith J D and Toms D 1996 {\it Phys. Rev. D 53 5771}\\
Titchmarsh E C 1948 {\it Introduction to the Theory of Fourier Integrals}
(London:Oxford University Press)\\
Toms D J 1992 {\it Phys. Rev. Lett. 69 1152}\\
{---------} 1993 {\it Phys. Rev. D 47 2483}\\
Trucks M 1998 {\it Phys. Lett. B 445 117}

\end{document}